# Multimode Optical Fiber Transmission with a Deep Learning Network

Babak Rahmani,[1,*] Damien Loterie,[1] Georgia Konstantinou,[1] Demetri Psaltis,[2] Christophe Moser[1]

[1]*Ecole Polytechnique Fédérale de Lausanne, Laboratory of Applied Photonics Devices, CH-1015 Lausanne, Switzerland*

[2]*Ecole Polytechnique Fédérale de Lausanne, Laboratory of Optics, CH-1015 Lausanne, Switzerland*

*Corresponding author: [babak.rahmani@epfl.ch](babak.rahmani@epfl.ch)

Multimode fibers (MMF) are an example of a highly scattering medium which scramble the coherent light propagating within them and produce seemingly random patterns. Thus, for applications such as imaging and image projection through a MMF, careful measurements of the relationship between inputs and outputs of the fiber are required. We show, as a proof of concept, that a deep learning neural network can learn the input-output relationship in a 0.75 m long MMF. Specifically, we demonstrate that a deep convolutional neural network (CNN) can learn the non-linear relationships between the amplitude of the speckle pattern obtained at the output of the fiber and the phase or amplitude at the input of the fiber. Effectively the network performs a non-linear inversion task. We obtained image fidelity (correlation) of ~98% compared with the image obtained using the measured matrix of the system. We further show that the network can be trained for transfer learning, i.e. it can transmit images through the MMF which belongs to another class which were not used for training/testing.

**Keywords:** multimode fibers; neural networks; deep learning; image transmission;

# 1. Introduction

Multimode fibers (MMF) scramble the waves propagating inside them and produce seemingly random patterns known as speckles at their outputs. Despite this seemingly random nature, the system consisting of an input pattern, propagation through a MMF and a detector, behaves deterministically and linearly when the optical power is below a certain threshold. It has been shown by many, that image transmission or imaging through a MMF, could be carried out, for example, by analog phase conjugation[1-5], by digital iterative methods[6-10], digital phase conjugation[11,12] or measuring experimentally the amplitude and phase of the output patterns corresponding to each input pattern to construct a matrix of complex numbers relaying the input to the output[13-16]. When the launched optical power in the fiber increases, non-linear effects start changing the way light propagates in the fiber and the system no longer remains linear and hence the matrix model becomes less and less accurate as power increases, degrading the imaging performance[17]. We explore, in this paper, an alternative to the matrix approach to describe the input-output relationship in a MMF which is based on learning the relation between the input and output without a priori knowledge of light propagation in the MMF system.

The idea of using neural networks in conjunction with MMFs has been around for almost three decades[18-20]. In Ref. (18), a neural network with a three layer perceptron structure was used to classify 10 categories of images transmitted through the fiber. This simple neural network could not recognize images for which it was not trained for. In another experiment with a MMF, a single hidden layer neural network was utilized to classify a speckle pattern corresponding to an input code for the purpose of increasing the transmission capacity of the fiber optic system[19,20].

Today, thanks to the ubiquitous availability of processing power via graphical processing units and new type of neural network architectures, a revival of applications using neural networks is happening and summarized in Ref. (21). The convolutional neural network (CNN) is a sub-class of neural networks which has been proposed to surpass the performance of other neural networks by decreasing the computation cost of fully connected layers through parameter sharing and use of sparse filters while, at the same time, increasing the number of layers in the network to achieve deep networks for solving more complex problems and speeding up the computations. With this new computational power, CNNs have been recently applied to imaging systems[22]. For example, in microscopy, deep convolutional neural networks (CNN) have been successfully used to provide resolution enhancement in images of a same class of histology samples[23], provide phase recovery in inverse problems[24,25] (obtaining the phase at one plane when the intensity is measured at another plane).

In this work, we propose to use deep CNNs to learn the propagation of light in a MMF. Our ultimate motivation for this work is linked to finding a method to control the propagation of light in a MMF when non-linear effects are present, for example in high power light transmission. In this case, the propagation of light is difficult to describe physically and cannot be modeled as a transmission matrix. We have for example observed that the enhancement of a focused spot at the output of a multi-core fiber degrades when the light energy is above 0.5 uJ (for a 500 fs input pulse duration)[17].

By a combination of experiments and computations, as explained below, we demonstrate, as a proof of concept, that a deep CNN is able to learn the transmission of light in a MMF in the linear regime (continuous wave). We project phase patterns on a phase only spatial light modulator, which are then sent into a MMF. At its output, we record the output speckle holographically with an off-axis coherent reference wave on a 2D camera. The holograms have information about the phase at the output of the MMF.

In a first part, we send successively plane waves at different angles as inputs to the fiber, with the appropriate phase mask on the SLM, and record their speckle holograms at the output. In this way, the transmission matrix of the system is measured and its inverse calculated[16]. In what follows, we use the transmission matrix to propagate the optical fields from the input to the output of the MMF and vice-versa. The transmission matrix method has been shown to be a robust method to characterize accurately the transmission of light through MMFs and thus provide a convenient method to perform the "virtual" experiments explained below.

A CNN is trained in two ways: by feeding 1. The amplitude of the speckle pattern (no phase information) and 2. The raw speckle holograms containing the phase information as inputs to the network. The corresponding outputs of the network are the phase patterns of the SLM.

We report that after training with 80% of the data, the CNN is able to reproduce the input SLM phase patterns with 2D correlations for the test set as high as 97.28% and 98.25% for the hologram and amplitude case respectively. Essentially, the CNN was set-up to learn the reverse propagation from fiber output to fiber input. Although the correlation of the computed SLM phase patterns by the CNN with the original inputs is high and the total set of input-output patterns is enough to measure fully the system transmission matrix in the Fourier basis, we observe that when an arbitrary (unseen) pattern is presented at the input of the network, it is unable to produce the SLM phase pattern which yields the proper output.

We hypothesize that the Fourier basis used for measuring the transmission matrix of the MMF is not an appropriate "basis" to train the CNN.

Hence, in the second part of this work, we train the CNN with a set of inputs taken from the handwritten Latin alphabet. Each image in this set is richer in terms of input frequency components than an image of the Fourier basis (cosines). We show that images belonging to a different class (handwritten digits) and which were not used in the training (and test) were able to be transmitted with ~90% fidelity through the MM fiber. As mentioned above, the system transmission was done computationally using the experimentally measured matrix. This result shows that the trained CNN was able to do transfer learning. To our knowledge, this has not been demonstrated before in MMFs.

## 2. Methods

*Experimental set-up:* The data for training the system is obtained with the setup depicted schematically in Fig. 1. The system here is a step-index (length=0.75 m) MMF with 50um diameter silica core and numerical aperture 0.22 (1055 number of fiber modes[26]). The inputs correspond to 2D phase patterns displayed on a phase only spatial light modulator (SLM) which are then demagnified on the MMF entrance facet by the 4F system composed of lens L1 and OBJ1. The MMF output facet is imaged onto a camera and which interferes with a coherent beam (fiber split to SMF2) to yield an off-axis hologram.

The light source is a continuous wave source at 532nm and power 100mW. However, it is attenuated with a variable attenuator and only 1mW is used for the acquisition of the images. The light source is coupled into a single mode fiber. A fiber coupler is used for obtaining two different ports (SMF1 and SMF2). The light beam which is coming out of the SMF1 (object beam), is filtered by the polarizer LP1, collimated by the lens L4 and directed on the SLM. The pattern created by the SLM is imaged through the relay system (lens L1 and objective lens OBJ1) at the MMF input. The quarter wave plate (QWP1) before the fiber input changes the polarization from linear to circular (this polarization is better preserved in step-index fibers[27]). Then it travels through the fiber and at the output an identical relay system (OBJ2 and L2) magnifies the image of the output and projects it on the camera plane (the QWP2 converts the circular polarization back to linear).

The light beam originating from the SMF2 (reference beam) is filtered by the polarizer LP2 and collimated by the L5. The mirrors KM1, KM2 and the beam splitter BS are used to adjust the off-axis angle of the reference arm. The BS is used to combine both beams in order to interfere on the camera plane. The FC1 and FC2 are fiber clamps and the FH3 is a filter holder.

*The neural network:* The architecture of the CNN is schematically shown in Fig. 2. It consists of 12 layers. The first and the last ones are the encoder and the decoder, respectively, and the middle 10 layers constitute what is known as hidden layers in the neural networks terminology. The input layer maps the gray-scale input images (one channel) to 64 channels (stack of processed images) via a trainable convolutional unit. The hidden layer of the CNN is comprised of 10 layers. Each block in Fig. 2 represents one layer which is formed by two convolutional entities which are individually followed by an element of rectified linear unit (RELU) with mapping functionality $RELU(x) = \max(0, x)$. The convolutional units take the convolution of the input images $X_k$ using weights $W_k^i$, biases $B_k^i$ and complying with the formula $Conv_{W_k^i}(X_k) + B_k^i$ where the subscript $k$ indicates the layer number and the superscript $i = \{1, 2\}$ corresponds to the first and the second convolution operation in each and every layer. At the output of each layer, an additional max-pooling[28] unit is considered to decrease the widths and heights of the images passed through the convolutional filters by a factor of two. The final layer is also a convolutional layer which simply decodes the images from 64 channels back to the original single-channel images. Once, the images are obtained at the final layer of the CNN (feed-forward step), they are compared against their corresponding labels in a mean-squared error (MSE) sense. The cost function reads as

$$MSE(\theta) = \frac{1}{N \times M \times M} \sum_{i=1}^{M} \sum_{j=1}^{M} |Y_{i,j}^R(\theta) - Y_{i,j}^L|^2 \qquad (1)$$

where $\theta$ is the CNN's set parameters (including weights and biases ), $i$ and $j$ are the indices of the neural network reconstructed image $Y^R$ and label image $Y^L$, $M$ is the width and height of the images and $N$ is the mini-batch size. Once the cost function is calculated, it is optimized using Adaptive Moment Estimation optimization (ADAM) algorithm[29]. To obtain accurate results within reasonable time, we empirically choose the learning rate parameter of $10^{-4}$ in the optimization algorithm.

The program was written using Tensorflow, a Python-based open-source library developed by Google for implementing neural networks. The Tensorflow version used here is 1.5 running on Python version 3.5.4. The machine used to run the simulations is a windows-based computer with Intel (R) Xeon (R) CPU E5-2609 v3@ 1.9GHz and 32 GB of RAM. We used NVIDIA Quadro M4000 graphic processing unit (GPU) as the platform for running Tensorflow.

A sequence of phase patterns $\Phi_n(x,y)$, where index $n$ represents the numbering of the pattern in the sequence and $x, y$ denote the Cartesian variables, is first created and then loaded onto the SLM which is then imaged onto the fiber input facet. The speckle patterns formed due to the modal interference inside the fiber are collected at its output facet by the camera as raw off axis holograms. A preprocessing step is carried out to obtain the optical field amplitudes $O_n(x,y)$. First, the raw holograms are Fourier transformed and then all areas except the holographic real-order are masked out in the Fourier domain. By taking the inverse Fourier transform, the field amplitude is reconstructed. This process is illustrated in Fig. 3. Hence, by cropping the Fourier-domain image around the holographic real-order, the size of the images used for training are decreased from 576x576 to 51x51 pixels, speeding up the training procedure substantially.

The field amplitude at the output of the MMF is used as the input of the neural network and the corresponding phase pattern is used as the output label to compute the MSE. These input phase patterns are in fact phase gratings which yield planes waves by diffraction from SLM. The number of possible input plane waves filling the numerical aperture of the fiber is 1961. We measure the transmission matrix with the 1961 patterns, 50 times, with 10 minutes interval between measurements.

To speed up the network training step, only 100 angles are selected from each matrix measurement. In this way, 50 sets of data, each containing 100 input-output images of size $51 \times 51$ pixels are obtained from the transmission matrix. So, in total, there are 5000 input phase patterns and 5000 fiber output amplitude pattern. Out of these sets, 40 of them are randomly selected to be used as our training set, while the remaining 10 are used for validation.

### 3. Results

Figure 4 plots (a) the MSE as well as (b) 2D-correlation between the labels (c) and the reconstructed phase patterns (d) versus the number of iterations for both the training and the validation dataset. The training time of the neural networks in this case is ~1hr, 20min. We observe a correlation of 98.25% between the labels and the reconstructed phases on the validation set.

We also verified that by using all 1961 input-output images measured 10 times, and then training the network with 80% of the data, a fidelity of 76% is achieved on the validation dataset. By comparison, the training here takes ~14hrs.

The neural network was also trained with the raw holograms intensity (instead of the field amplitude) obtained at the output of the MMF. From a total of 50 sets each containing 100 images obtained separately from previous, 1000 are randomly selected for validation and 4000 for the training. (Raw holograms were resized from 576x576 to 51x51 to feed the network). Figure 5 depicts (a) the MSE as well as (b) 2D-correlation between the labels and the reconstructed input phase patterns versus the number of iterations for both the training and the validation dataset. The training time of the neural networks in this case is ~2hr, 30min. The time for training the network when raw holograms are fed is higher than the time for training with the optical field amplitudes. It is expected since for the case of raw holograms, there is the additional phase information between the propagating modes which have to be learned. Nevertheless, the performance of the network to learn the inverse propagation (output-input relation of the MMF system) is very similar in terms of correlation (98.25% vs. 97.28%).

Next, we choose a MMF input image (a cross picture) for which the corresponding output speckle does not belong to the category produced by the plane wave phase patterns. We observe in Fig. 6 that the CNN generates a phase pattern which is not similar to the desired input. Although the network has learned the input-output relationship with patterns representing the Fourier basis, it is not capable to effectively learn the generalized fiber transmission.

Interestingly, we show below that by changing the basis class of input-output images, we were able to train a CNN capable of reconstructing the heart picture, as an example, and other images belonging to a different class through the MMF from the amplitude speckled fiber output. We use images of the Latin alphabet adopted from Ref. (30) for training the network, and a digit dataset[30] for evaluating transfer learning reconstruction performance. The two dataset are thus not of the same class of images. We then use the transmission matrix obtained experimentally to obtain the speckle pattern amplitudes of alphabets/digits projected as input intensities/phases on the fiber. The neural network is then trained on the amplitude of the speckle patterns and input amplitude/phases of only the alphabet. Fig. 7 illustrates the MSE as well as 2D-correlation between the labels and the reconstructed SLM amplitude/phase patterns belonging to the Latin alphabet when the amplitude input image is used (Fig. 7 a, b) and when the phase input image is used (Fig. 7 c, d). Examples of the label and reconstructed amplitude/phase patterns in the validation dataset are shown in Fig. 7 (e, f).

Examples of (a) reconstructed amplitude input pattern and (b) reconstructed phase input pattern by the CNN belonging to another class are depicted in Fig 8. Example of reconstructed amplitude input pattern for a heart picture is also shown in (c). This demonstration shows the ability of the network to perform transfer learning. It should be further emphasized that the network here is in fact learning a non-linear inversion, i.e. inferring the input amplitude/phase from only the output amplitude and not the complex output (phase and amplitude)

### 4. Conclusion

In summary, the main result of our study is that by using a proper set of input–output images to a multimode fiber system, a deep convolutional neural network was able to be trained to achieve transfer learning i.e. transmit desired patterns which did not belong to the class of images used to train the network. Specifically, we found that the set of patterns belonging to the handwritten Latin Alphabet could train a 12 layers CNN which was then able to transmit the class of handwritten digits with ~90% fidelity. We used the amplitude as input to the network and the SLM amplitude/phase as output. It is noteworthy to realize that the network performs effectively a non-linear inversion task. The experiments were carried out by experimentally measuring the transmission matrix of the MMF system using the Fourier basis as inputs. Then the experimental matrix was used to transmit computationally the fields from the fiber input to the fiber output. This work was performed at optical intensities too low to create non-linear effects. These results could potentially be extended to learn image transmission when complex non-linear phenomena occur in MMF.

### Acknowledgments

Funding: This project was partially conducted with funding from the Swiss National Science Foundation (SNSF) project MuxWave (200021_160113\1).

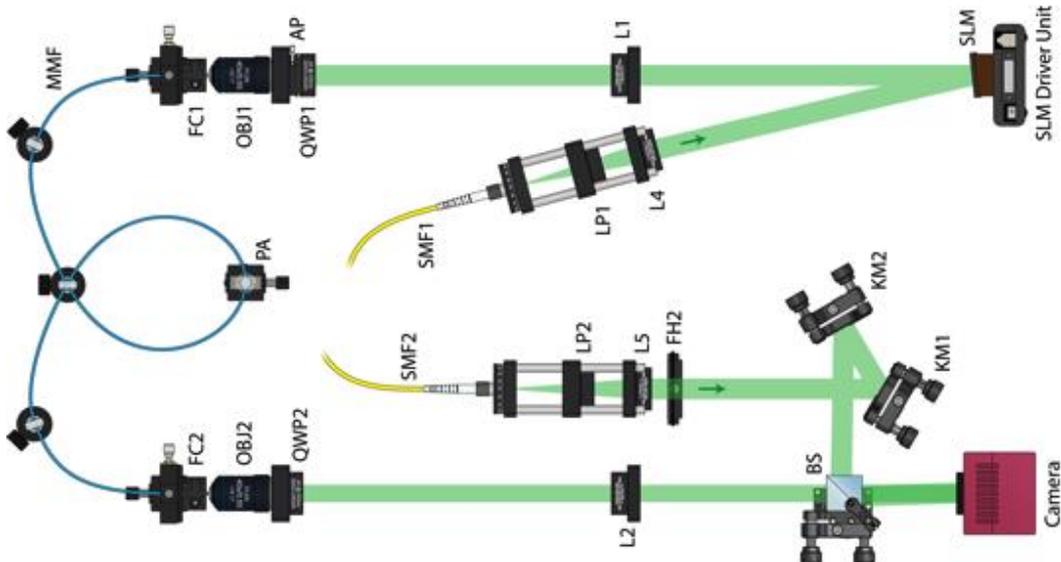

**Figure 1.** A schematic of the experimental setup used to obtain the input-output (speckle-phase patterns) for training the neural network.

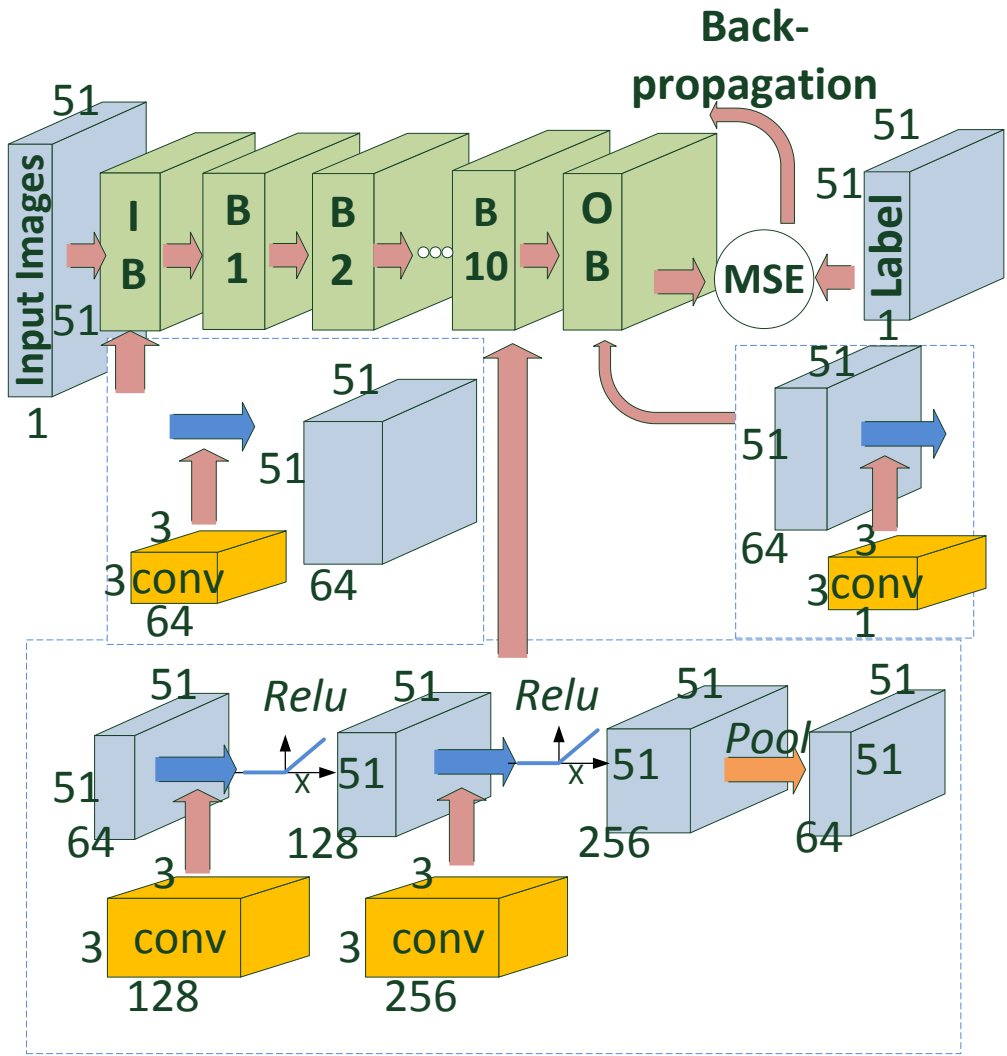

**Figure 2.** Detailed schematic of the CNN employed for training and testing. IB (Input Block), OB (Output Block), B$i$ (Block $i$ where $i$=1, 2,…, 10), Pool (Max pooling). The input block maps the input images via 64 convolutional filters. Each block in hidden layer contains two convolution layers followed by a max

pooling layer which down samples the widths and heights of the images by a factor of two. A rectified linear unit (Relu) transform is placed after each convolution unit in the hidden layer. The images are then mapped to the output channel via convolution filters in the output block. The mean squared error (MSE) between the label and processed images is then calculated and back propagated to the network to update the learnable variables.

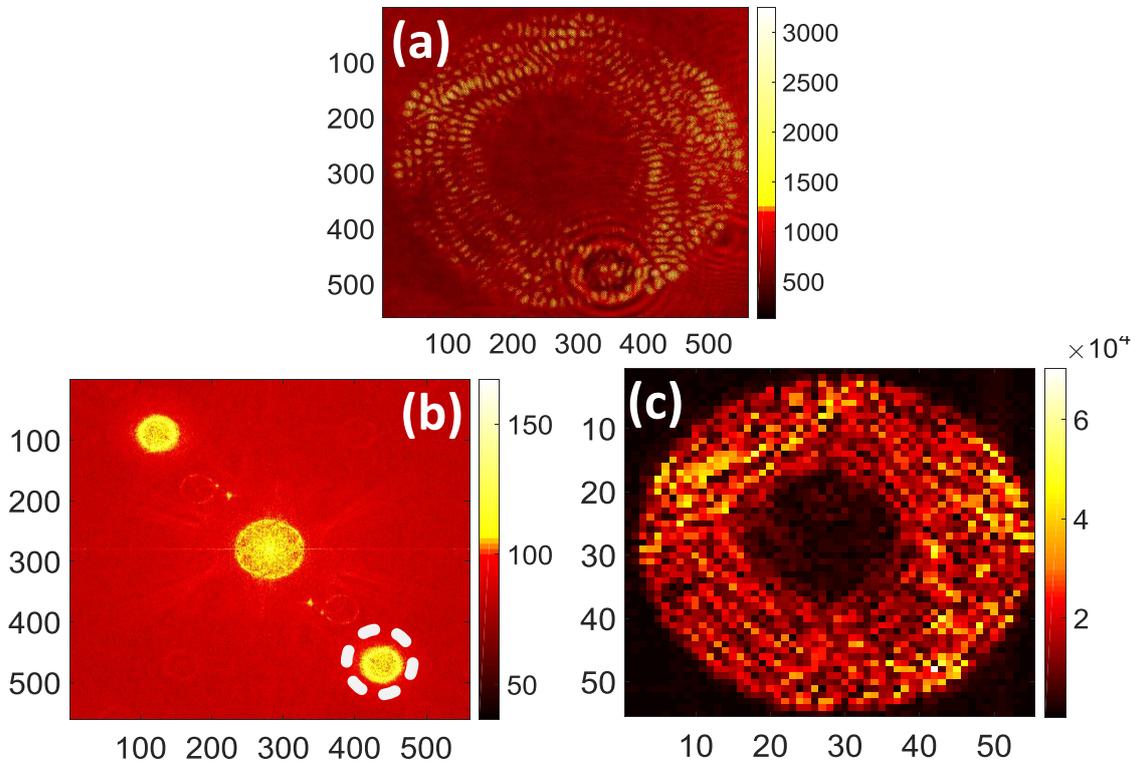

**Figure 3.** Object optical field amplitudes reconstruction steps. Raw holograms are first Fourier transformed and masked with proper filter. Inverse Fourier transform gives the reconstructed object wave. (a) Raw off-axis hologram at the output of the MMF as recorded by the camera. (b) Fourier transform of the hologram shown in (a), represented here with a logarithmic amplitude coloring. (c) inverse Fourier transform of masked out area (dashed circle in (b)).

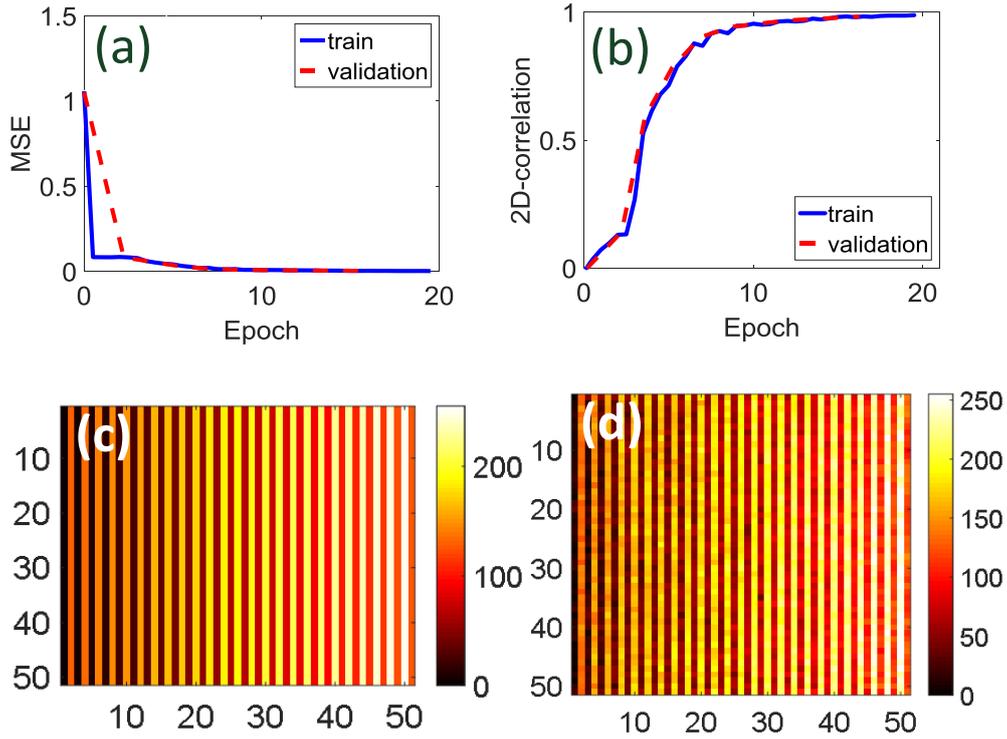

**Figure 4.** The performance of the neural network on the training and validation dataset. Calculated (a) MSE, (b) 2D-correlation between the labels (c), original phase pattern, and corresponding reconstructed phase patterns (d) by the CNN when the field amplitude of the MMF output patterns are fed to the network as input.

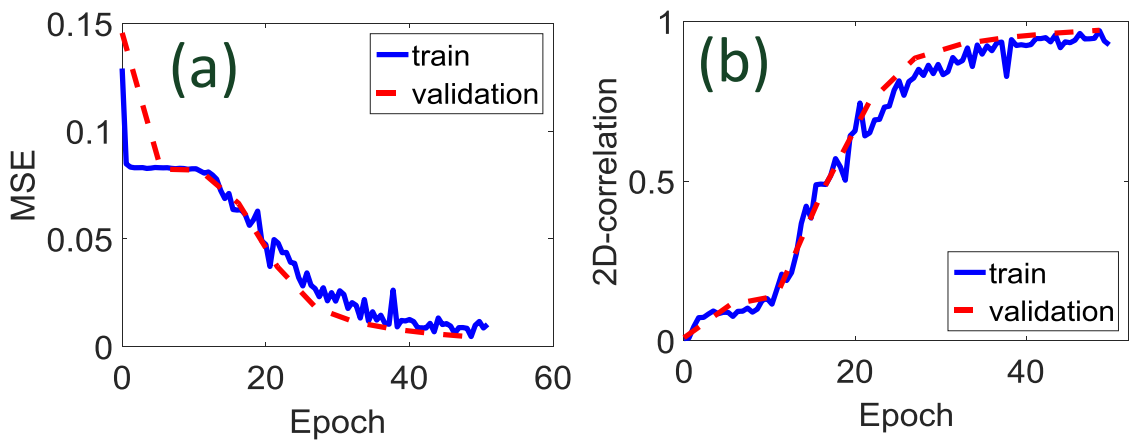

**Figure 5.** The performance of the neural network on the training and validation dataset. Calculated (a) MSE, (b) 2D-correlation between the labels and the reconstructed input phase patterns when raw holograms are fed to the CNN.

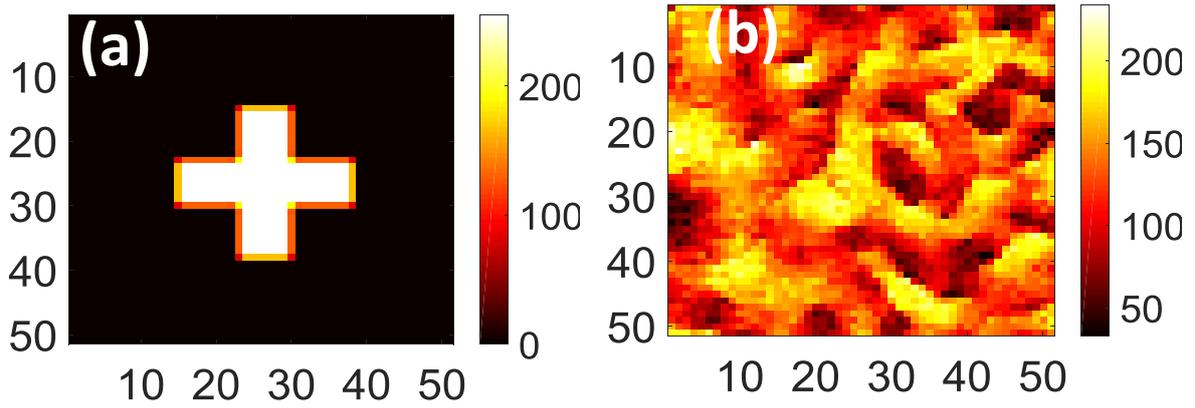

**Figure 6.** (a) Input phase image on the SLM from a category of image not used to train the CNN (b) computed phase image on the SLM by the CNN. The network is not able to reconstruct the proper pattern.

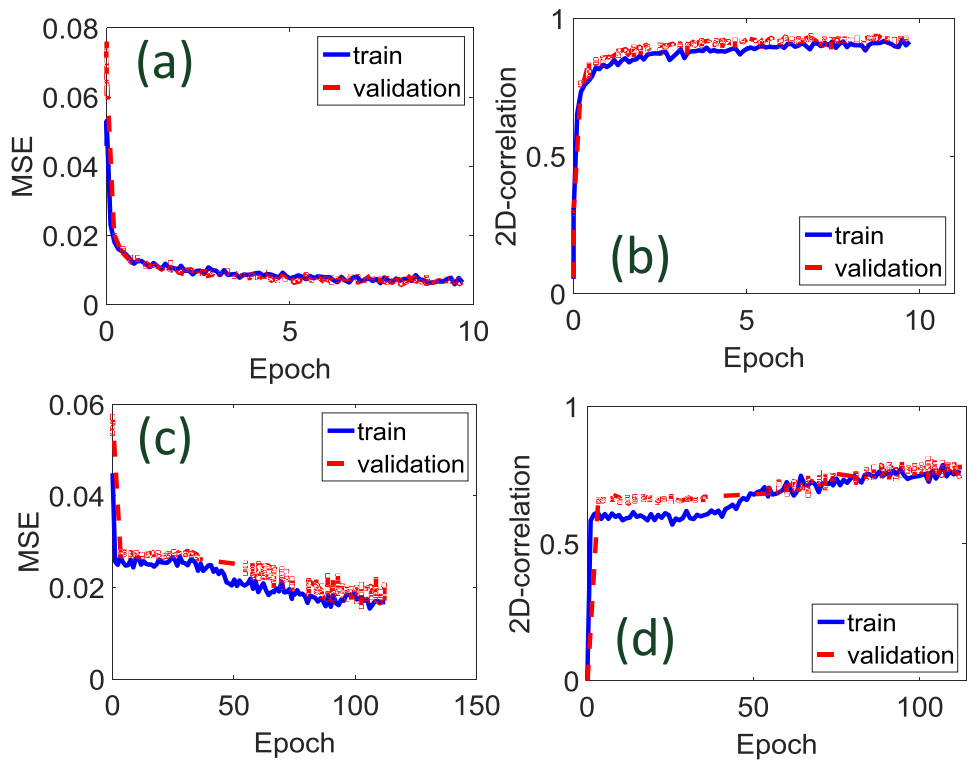

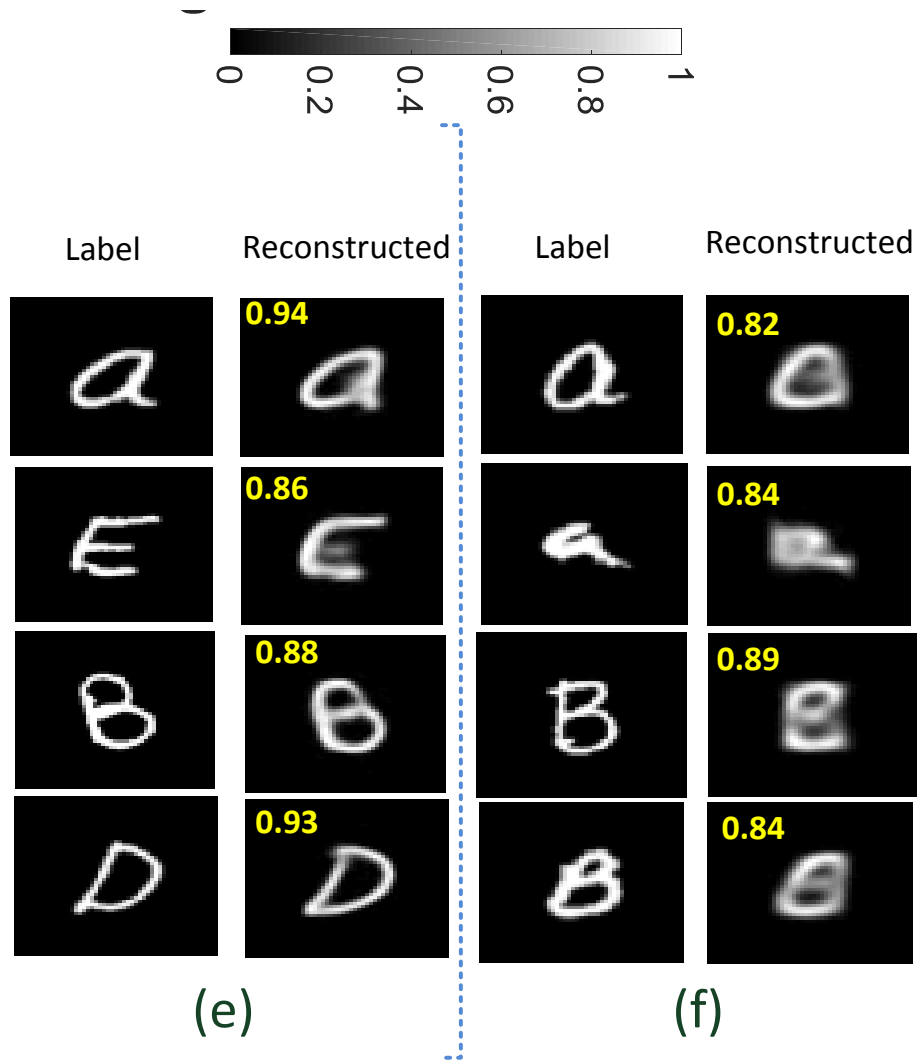

**Figure 7.** The performance of the neural network in learning the inverse transmission matrix when it is trained with the handwritten Latin Alphabet. Calculated (a) MSE, (b) 2D-correlation when amplitude input of Alphabets is used. Calculated (c) MSE and (d) 2D-correlation when phase input of Alphabets is used. Examples of (e) reconstructed amplitude input pattern and (f) reconstructed phase input pattern by the CNN. The fidelity number for each reconstructed image with respect to its corresponding label is shown.

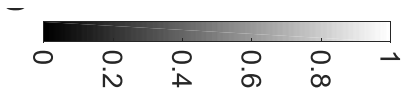

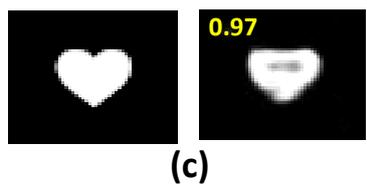

(a)　　　　　　　(b)

(c)

**Figure 8.** Performance of the neural network in transfer learning. (a) reconstructed amplitude input pattern and (b) reconstructed phase input pattern of digits. (c) Example of reconstructed amplitude input pattern for a heart picture. The fidelity for each reconstructed image with respect to its corresponding label is shown.